\documentclass[twocolumn,longbibliography, superscriptaddress, english, prx]{revtex4-1}
\usepackage[colorlinks=true,urlcolor=blue,citecolor=blue,linkcolor=blue]{hyperref} 
\usepackage[T1]{fontenc}
\usepackage[latin9]{inputenc}
\usepackage{amssymb}
\usepackage{graphicx}
\usepackage{amsmath,color}
\usepackage{mathrsfs}
\usepackage{float}
\usepackage{indentfirst}
\usepackage{txfonts}
\usepackage{mathtools}
\usepackage{comment}

\DeclarePairedDelimiterX{\infdivx}[2]{(}{)}{%
  #1\;\delimsize\|\;#2%
}
\newcommand{\KL}{\mathbb{KL}\infdivx}
\usepackage{algpseudocode} 
\usepackage{algorithm}

\makeatletter

\def\journal #1, #2, #3, 1#4#5#6{{\sl #1~}{\bf #2}, #3 (1#4#5#6) }

\def\eqa{\begin{eqnarray}}
\def\eea{\end{eqnarray}}
\newcommand{\eq}{\begin{equation}}
\newcommand{\ee}{\end{equation}}

\newcommand{\Eq}[1]{Eq.~(\ref{#1})}

\DeclareMathOperator*{\E}{\mathbb{E}}

\makeatother

\usepackage{babel}

\begin{document}

%\title{Metropolis independent sampler using invertible flow}
%\title{Invertible Variational Renormalization Group Flow Using Probability Density Distillation}
%\title{Invertible Variational Renormalization Group Flow via Deep Neural Networks}
%\title{Variational Renormalization Group Flow}
%\title{Variational Renormalization Group Flow with Invertible Deep Neural Networks}
%\title{Bijective Variational Renormalization Group Flow with Deep Neural Nets}
%\title{Deep Learning Bijective Variational Renormalization Group Flow}
%\title{Deep Learning Variational Renormalization Group}
\title{Neural Network Renormalization Group}
\author{Shuo-Hui Li}
\affiliation{Institute of Physics, Chinese Academy of Sciences, Beijing 100190, China}
\affiliation{University of Chinese Academy of Sciences, Beijing 100049, China}
\author{Lei Wang}
\email{wanglei@iphy.ac.cn}
\affiliation{Institute of Physics, Chinese Academy of Sciences, Beijing 100190, China}
\affiliation{Songshan Lake Materials Laboratory, Dongguan, Guangdong 523808, China} 

\begin{abstract}
We present a variational renormalization group (RG) approach based on a reversible generative model with hierarchical architecture. The model performs hierarchical change-of-variables transformations from the physical space to a latent space with reduced mutual information. Conversely, the neural network directly maps independent Gaussian noises to physical configurations following the inverse RG flow. The model has an exact and tractable likelihood, which allows unbiased training and direct access to the renormalized energy function of the latent variables. To train the model, we employ probability density distillation for the bare energy function of the physical problem, in which the training loss provides a variational upper bound of the physical free energy. We demonstrate practical usage of the approach by identifying mutually independent collective variables of the Ising model and performing accelerated hybrid Monte Carlo sampling in the latent space. Lastly, we comment on the connection of the present approach to the wavelet formulation of RG and the modern pursuit of information preserving RG.  
%The approach could be useful for automatically identifying collective variables and effective field theories.
\end{abstract}
\maketitle

%\section{Introduction}
%Deep Learning and RG 
The Renormalization group (RG) is one of the central schemes in theoretical physics, whose impacts span from high-energy~\cite{Gell-Mann1954} to condensed matter physics~\cite{Wilson1971a,Wilson1971b}. In essence, RG keeps the relevant information while reducing the dimensionality of statistical data. Besides its conceptual importance, practical RG calculations have played important roles in solving challenging problems in statistical and quantum physics~\cite{Wilson1975, Swendsen1979}. A notable recent development is to perform RG calculations using tensor network machinery~\cite{Vidal2006,Levin2007,Gu2008, Gu2009, Xie2009, Zhao2010,Xie2012,Efrati2014,Evenbly2015a,Evenbly2015b, PhysRevLett.118.110504, Bal2017, PhysRevB.97.045111}

The relevance of RG goes beyond physics. For example, in deep learning applications, the inference process in image recognition resembles the RG flow from microscopic pixels to categorical labels. Indeed, a successfully trained neural network extracts a hierarchy of increasingly higher-level concepts in its deeper layers~\cite{zeiler2014visualizing}. In light of such intriguing similarities, Refs~\cite{Beny2013, Mehta2014, Bradde2017, Iso2018} drew connections between deep learning and the RG, Ref.~\cite{Koch-Janusz2017} proposed an RG scheme based on mutual information maximization, Ref.~\cite{You2017} employed deep learning to study holography duality,
%and Refs.~\cite{Carrasquilla2017, VanNieuwenburg2017, Wang2016} investigated phase transitions from the machine learning perspective. 
and Ref.~\cite{Kenway} examined the adversarial examples from a RG perspective. Since the discussions are not totally uncontroversial~\cite{Mehta2014, Lin2017, Schwab2016, Koch-Janusz2017, Iso2018}, 
%Indeed, the downsampling steps in the convolutional neural network (CNN) are similar to the block spin renormalization group~\cite{blockspinRG}. 
%Despite these discussions, it is remains unclear what is the precise connections between deep learning and RG. 
it remains highly desirable to establish a more concrete, rigorous, and constructive connection between RG and deep learning. Such a connection will not only bring powerful deep learning techniques into solving complex physics problems but also benefit theoretical understanding of deep learning from a physics perspective.  

In this paper, we present a neural network based variational RG approach (NeuralRG) for statistical physics problems. In this scheme, the RG flow arises from iterative probability transformation in a neural network. Integrating latest advances in deep learning including \emph{Normalizing Flows}~\cite{Dinh2014, Germain2015, Rezende2015, Dinh2016, Kingma2016, Oord2016c, Oord2016a, Papamakarios2017, Diederik2018} and \emph{Probability Density Distillation}~\cite{Oord2017}, and tensor network architectures, in particular, the multi-scale entanglement renormalization ansatz (MERA)~\cite{Vidal2006}, the proposed NeuralRG approach has a number of interesting theoretical properties (variational, exact and tractable likelihood, principled structure design via information theory) and high computational efficiency. 
%We anticipate applying of this approach to more challenging problems in the statistical physics problems.
%\section{Network Architecture}
%Transformation of the probability is an old idea in the simulation. Conventially, it is however very expensive to find such transformations. Invertible flow parametizes a family of networks that performs invertible transformations of the prior probability distribution. 
The NeuralRG approach is closer in spirit to the original proposal based on Bayesian net~\cite{Beny2013} than more recent discussions on Boltzmann Machines~\cite{Mehta2014, Iso2018} and Principal Component Analysis~\cite{Bradde2017}. 

\begin{figure}[t]
\centering
\includegraphics[width=\columnwidth]{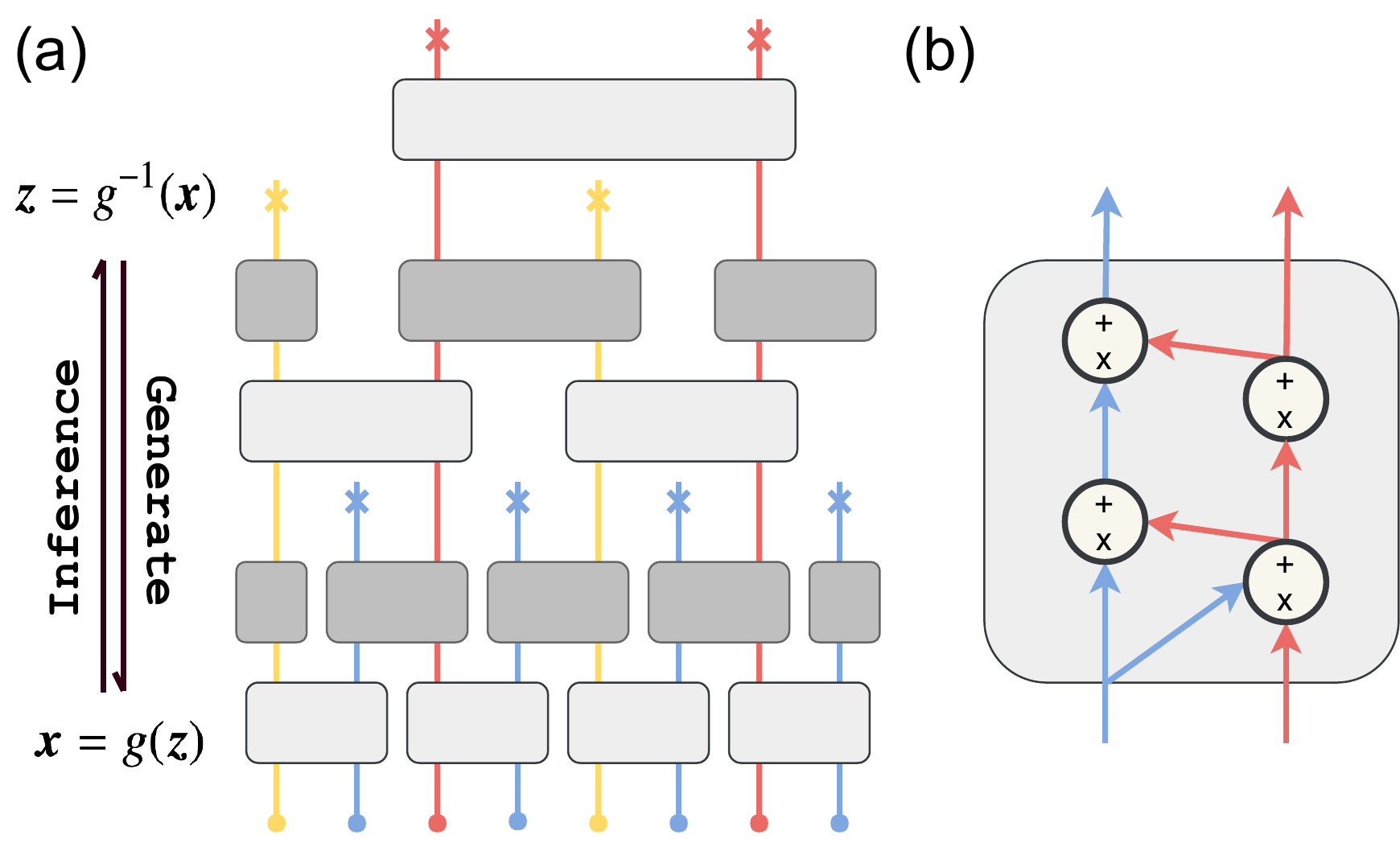}
\caption{(a) The NeuralRG network is formed by stacking bijector networks into a 
hierarchical structure. The solid dots at the bottom are the physical variables $\boldsymbol{x}$ and the crosses are the latent variables $\boldsymbol{z}$. 
%The stars denote the renormalized collective variables at various scales. 
Each block is a bijector. The light gray and the dark gray blocks are the disentanglers and the decimators respectively. The RG flows from bottom to top, which corresponds to the inference of the latent variables conditioned on the physical variables. Conversely, one can directly generate  physical configurations by sampling the latent variables according to the prior distribution and passing them downwards through the network. (b) The internal structure of the bijector block consists of Normalizing Flows~\cite{Dinh2016}.}
\label{fig:nflow}
\end{figure}

%Neural RG architecture 
Figure~\ref{fig:nflow}(a) shows the proposed architecture. Each building block is a diffeomorphism, i.e., a bijective and differentiable function parametrized by a neural network, denoted by a bijector~\cite{Dillon2017, PyroDevelopers2017}. Figure~\ref{fig:nflow}(b) illustrates one realization of the bijector using real-valued nonvolume preserving flows (Real NVP)~\cite{Dinh2016}~\cite{SM}, which is one of the reversible generative models known as the Normalizing Flows~\cite{Dinh2014, Germain2015, Rezende2015, Dinh2016, Kingma2016, Oord2016c, Oord2016a, Papamakarios2017,Diederik2018}. 

The network relates the physical variables $\boldsymbol{x}$ and the latent variables $\boldsymbol{z}$ via an invertible transformation $\boldsymbol{x} = g(\boldsymbol{z})$. Their probability densities are also related~\cite{Goodfellow2016} 
\begin{equation}
%q(\boldsymbol{x}) = p (\boldsymbol{z}) \left| \det\left (\frac{\partial \boldsymbol{z}}{\partial \boldsymbol{x}}\right) \right|.
\ln q(\boldsymbol{x}) =\ln p (\boldsymbol{z}) - \ln \left| \det\left (\frac{\partial \boldsymbol{x}}{\partial \boldsymbol{z}}\right) \right|,
\label{eq:probflow}
\end{equation}
where $q(\boldsymbol{x})$ is the normalized probability density of the physical variables. And  $p(\boldsymbol{z})=\mathcal{N}(\boldsymbol{z};\boldsymbol{0},\boldsymbol{1})$ is the prior probability density of the latent variables chosen to be a normal distribution.  
%$p(\boldsymbol{z})= e^{-|\boldsymbol{z}|^{2}/2}/(2\pi)^{N/2}$. 
%We view the $ \boldsymbol{z}$ as latent variables with a
The second term of \Eq{eq:probflow} is the log-Jacobian determinant. %,which can be easily computed by collecting the contributions from each bijector block. 
Since the log-probability can be interpreted as a negative energy function, \Eq{eq:probflow} shows that the renormalization of the effective coupling is provided by the log-Jacobian at each transformation step. %The equation relates the physical variables and uncorrelated Gaussian latent variables.

Since diffeomorphisms form a group, an arbitrary composition of the building blocks is still a bijector. This motivates the modular design shown in Fig.~\ref{fig:nflow}(a). The layers alternate between disentangler blocks and decimator blocks. The disentangler blocks in light gray reduce correlation between the inputs and pass on less correlated outputs to the next layer. While the decimator blocks in dark gray pass only a subset of its outputs to the next layer and treat the remaining ones as irrelevant latent variables indicated by the crosses. %while the others still carries information about the input variables.
The RG flow corresponds to the inference of the latent variables given the physical variables, $\boldsymbol{z}=g^{-1}(\boldsymbol{x})$. The kept degrees of freedom emerge as renormalized collective variables at coarser scales during the inference.  
%renormalized collective degrees of freedom emerge during the transformation. %After training of the neural network, for each given physical configuration one can perform inference to obtain the corresponding latent variables. The kept variables act as renormalized variables at coarser levels.
%the neural network performs RG transformation where collective degrees of freedom may emerge automatically.
%While $\boldsymbol{x}$ as the physical variables which exhibit complex probability distribution of the target problem.
In the reversed direction, the latent variables are injected  into the neural network at different depths. And they affect the physical variables at different length scales. %Obtaining physical variables corresponds to direct ancestral sampling from top to bottom in the neural network. 

%One can sample the physical variables through a direct ancestral sampling by drawing latent variables according to the prior distribution $p(\boldsymbol{z})$ and passing them 
%The bijective property is crucial for learning the RG flow in a controlled way. %because one has  controlled probability transformation while still keeping tractability. 
%No matter how complex is the hierarchical transformations performed by the neural network, one can efficiently compute the normalized probability density $q(\boldsymbol{x})$ for either given or sampled physical configuration $\boldsymbol{x}$ by keeping track of the Jacobian determinant of each bijector locally. 

%adopt the filter used and train it for larger system size. 
%The blocks trained at smaller system size thus provide good initialization of the larger system size. %We can then perform fine tuning at the larger system size. 

%This bijective renormalization group idea is the transformation of probability density under invertible functions.  
%Assuming the are related by an invertible function $\boldsymbol{x} = f^{-1}(\boldsymbol{z}) \equiv g(\boldsymbol{z})$. 

%Connections to tensor network and quantum circuits
%TNS connection and difference
The proposed NeuralRG architecture shown in Fig.~\ref{fig:nflow}(a) is largely inspired by the  MERA structure~\cite{Vidal2006}. In particular, stacking bijectors to form a reversible transformation is analogous to the quantum circuit interpretation of MERA. The difference is that the neural network transforms probability densities instead of quantum states. 
%Exploiting these analogies provide constructive guidelines to neural network architecture design. %Despite of these similarity, the difference that one manipulates the probability density instead of the quantum states in this case. 
Compared to the tensor networks, the neural network has the flexibility that the blocks can be arbitrarily large and long-range connected. Moreover, arbitrary complex NeuralRG architecture constructed in a modular fashion can be trained efficiently using differentiable programming frameworks~\cite{Abadi2016, Paszke2017}. In practice, one can let the bijectors in the same layer share weights due to the translational invariances of the physical problem~\footnote{If needed, one can even share weights in the depth direction due to scale invariance emerged at criticality. The scale-invariant reduces the number of parameters to be independent of the system size. In this case, one can iterate the training process for increasingly larger system size and reuse the weights from the previous step as the initial value.}.

\begin{figure}[t]
\centering
\includegraphics[width=\columnwidth]{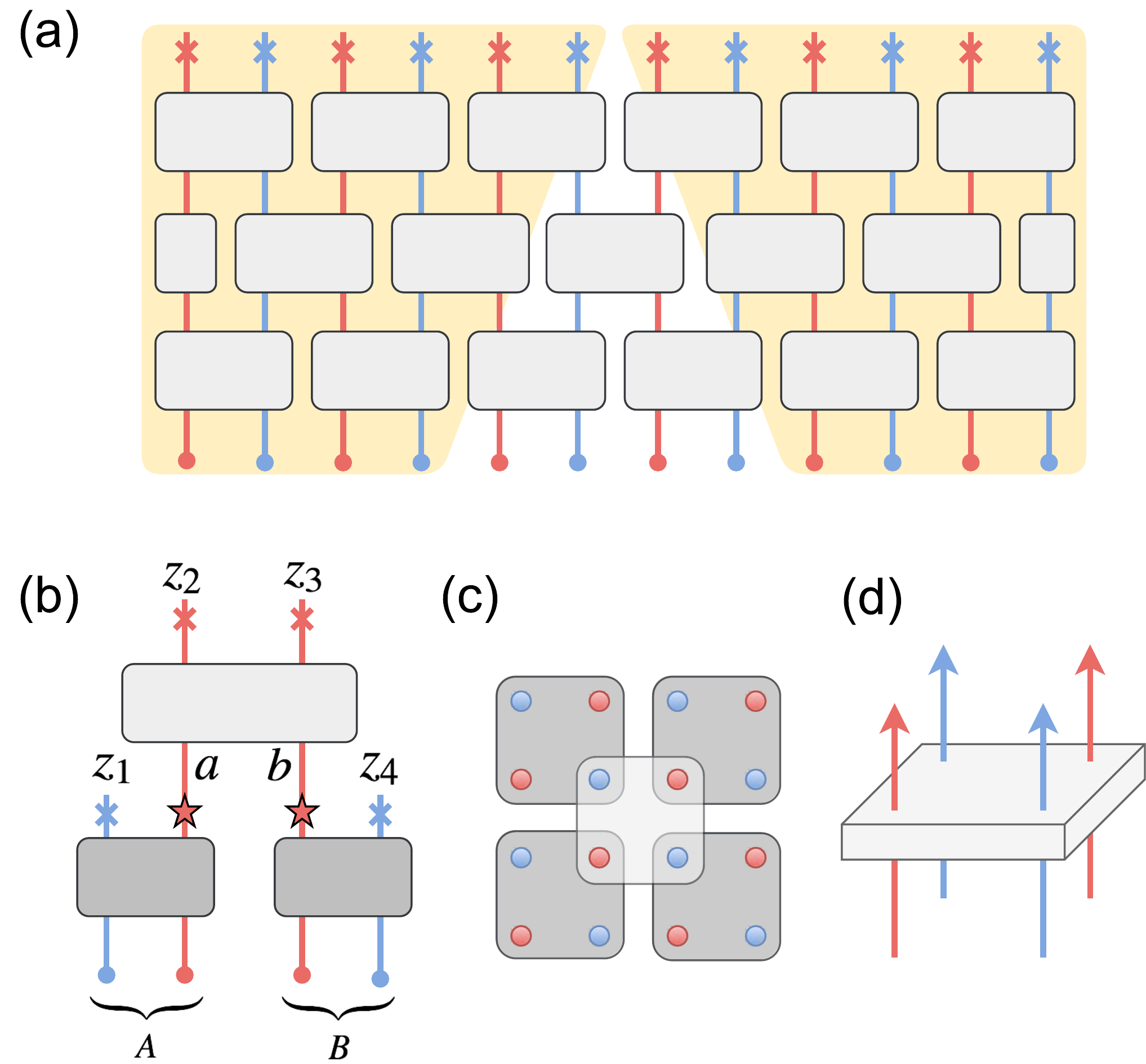}
\caption{(a) A reference neural network architecture with only disentanglers. The physical variables in the two shaded regions are uncorrelated because their causal light cones do not overlap in the latent space. (b) Mutual information is conserved at the decimation step, see \Eq{eq:MI}. (c) The arrangement of the bijectors in the two-dimensional space. (d) Each bijector acts on four variables. Disentanglers reduce mutual information between variables. While for decimators, only one of its outputs is passed on to the next layer and the others are treated as latent variables.}
\label{fig:tebd-tree-2d} 
\end{figure}

%TEBD and Tree
Compared to ordinary neural networks used in deep learning, the architecture in Fig~\ref{fig:nflow}(a) has stronger physical and information theoretical motivations. To see this, we consider a simpler reference structure shown in Fig.~\ref{fig:tebd-tree-2d}(a) where one  uses disentangler blocks at each layer. The resulting structure resembles a time-evolving block decimation network~\cite{Vidal2003}. Since each disentangler block connects only a few neighboring variables, the causal light cone of the physical variables at the bottom can only reach a region of latent variables proportional to the depth of the network. Therefore, the correlation length of the physical variables is limited by the depth of the disentangler layers. The structure of Fig.~\ref{fig:tebd-tree-2d}(a) is sufficient for physical problems with  finite correlation length, i.e. away from the criticality.  %Moreover, since all the variables are kept until the end, the TEBD-like network is not performing coarse-graining renormalization transformation 

On the other hand, a network formed only by the decimators is similar to the tree tensor network~\cite{Shi2006}. For example, the mutual information (MI) between the variables at each decimation step shown in Fig.~\ref{fig:tebd-tree-2d}(b) follows
\begin{equation}
I (A:B) = I (z_1\cup a  : b\cup z_4) = I (a:b).
\label{eq:MI} 
\end{equation}
The first equality is due to the MI being invariant under invertible transformation of variables within each group. While the second equality is due to the random variables $z_1$ and $z_4$ being independent of all other variables. Applying \Eq{eq:MI} recursively at each decimation step, one concludes that the MI between two sets of physical variables is limited by the top layer in a bijector net of the tree structure. 
One thus needs to allocate sufficient resources in the  bottleneck blocks to successfully capture the MI of the data. 
%This does not necessarily impose a constraint on the expressibility of the network since the MI between two continuous variables can be arbitrarily large in principle. However, a neural network with decimators only could be limited in practice since it is rather unphysical to carry the MI between two extensive regions with only two variables~\footnote{The decimator only tree structure NeuralRG network is sufficient to model one dimensional physical systems with short-range interactions since the MI is constant~\cite{Wolf2008}.}.

It is straightforward to generalize the NeuralRG architecture in Fig.~\ref{fig:nflow}  to handle data in higher dimensional space. For example, one can stack layers of bijectors in the form of Fig.~\ref{fig:tebd-tree-2d}(c). These  bijectors accept $2\times 2$ inputs as shown in Fig.~\ref{fig:tebd-tree-2d}(d). For the decimator, only one out of four outputs is passed on to the next layer. In a network with only disentanglers, the depth should scale linearly with system size to capture diverging correlation length at criticality. While the required depth only scales logarithmically with system size if one employs the MERA-like structure. Note that different from the tensor network modeling of quantum states~\cite{Barthel2010}, the MERA-like architecture is sufficient to model classical systems with short-range interactions even at criticality since they exhibit  the MI area law~\cite{Wolf2008}. 

%Learning
Building the neural network using Normalizing Flows provides a generative model with explicit and tractable likelihoods compared to previous studies~\cite{Mehta2014, Koch-Janusz2017, Liu2017b, Huang2017a, Liu2017a, Iso2018}. This feature is valuable for studying physical problems because one can have unbiased and quantitative control of the training and evaluation of the model.  
% provides direct access to the exact log-likelihood~\Eq{eq:probflow},
Consider a standard setup in statistical physics, where one has accesses to the bare energy function, i.e. the \emph{unnormalized} probability density ${\pi}(\boldsymbol{x})$ of a physical problem, direct sampling of the physical configurations is generally difficult due to the intractable partition function $ Z  = \int \mathrm{d}{\boldsymbol{x}}\, {\pi}(\boldsymbol{x})$%and exponentially small probability densities in high dimensions
~\cite{MacKay2005}.
The standard Markov chain Monte Carlo (MCMC) approach
%, in which only ratios between unnormalized probability densities are used to collect samples~\cite{Metropolis1953}. However, sampling using MCMC 
suffers from the slow mixing problem in many cases~\cite{MCBookJunLiu}. 
%Actually, in general it can be difficult problem to collect these training samples in the first place due to two reasons~\cite{MacKay2005}. First, one does not have access to the normalized probability distribution since the partition function $
%Z  = \int \mathrm{d}{\boldsymbol{x}}\,  {\pi}(\boldsymbol{x}), 
%Z  = \sum_{\boldsymbol{x}} e^{-E(\boldsymbol{x})}, 
%$ is in general intractable. Second, even if one knew the normalized density it is still unrealistic to sample from them directly in high dimensional space. 

%Although certain bootstrap approach for the application of to the physics problems~\cite{Liu2017, Song2017}.
% one typically has direct access the \emph{unnormalized} probability distribution ${\pi}(\boldsymbol{x})$ but do not  access data of the statistical physics model. 

%Consider a probability distribution of continuous variables $\pi(\boldsymbol{x})$, change of variables can be a very efficient approach in low dimensions~\cite{Devroye}. We aim to model the probability distribution using a normalized probability $q(\boldsymbol{x})$. 
%Logarithmic of the unnormalized probability is the energy function of the physical system. 

%We avoid the aforementioned problem 
%In light of the aforementioned issues, 
We train the NeuralRG network by minimizing the \emph{Probability Density Distillation} (PDD) loss 
\begin{equation}
\mathcal{L} = \int \mathrm{d}{\boldsymbol{x}}\,  q(\boldsymbol{x}) \left[ \ln{q(\boldsymbol{x})} - \ln{{\pi} (\boldsymbol{x}) } \right], 
%\mathcal{L} =\sum_{\boldsymbol{x}} q(\boldsymbol{x}) \left[ E(\boldsymbol{x})  + \ln{q(\boldsymbol{x})}  \right],
\label{eq:loss}  
\end{equation}
which was recently employed by DeepMind to train the Parallel WaveNet~\cite{Oord2017}. The first term of the loss is the negative entropy of the model density $q(\boldsymbol{x})$, which favors diversity in its samples. While the second term corresponds to the expected energy since $-\ln \pi(\boldsymbol{x})$ is the energy function of the physical problem. %This term increases the model probability density at those more probable configurations. 

%entropy of the model probability distribution and the cross energy with the target model. 
In fact, the loss function \Eq{eq:loss} has its origin in the variational approaches in statistical  mechanics~\cite{Feynman1972, MacKay2005,Bishop2006}. To see this, we write
\begin{equation} 
% \mathcal{L} \ge  - \ln   \sum_{\boldsymbol{x}} q(\boldsymbol{x})  \left[\frac{ e^{-E(\boldsymbol{x})} }{q(\boldsymbol{x})} \right] = -\ln Z. 
% \mathcal{L} \ge  - \ln   \sum_{\boldsymbol{x}} q(\boldsymbol{x})  \left[\frac{ \pi(\boldsymbol{x}) }{q(\boldsymbol{x})} \right] = -\ln Z. 
 \mathcal{L} +\ln Z = \KL*{q(\boldsymbol{x})}{\frac{\pi (\boldsymbol{x})}{Z}} \ge 0, 
\label{eq:lowerbound} 
\end{equation}
where the Kullback-Leibler (KL) divergence measures the proximity between the model and the target probability densities~\cite{Bishop2006, Goodfellow2016}. Equation (\ref{eq:lowerbound}) reaches zero only when the two distributions are identical. One thus concludes that the loss \Eq{eq:loss} provides a variational upper bound of the physical free energy of the system, $-\ln Z$. 
%In the This loss function is also known as the evidence lower bound in the machine learning literature~\cite{}. %Therefore, the probability density distillation pwocess can be understood as a variational calculation. 

For the actual optimization of the loss function, we randomly draw a batch latent variables according to the prior probability $p(\boldsymbol{z})$ and pass them through the generator network  $\boldsymbol{x} = g(\boldsymbol{z})$, an unbiased estimator of the loss \Eq{eq:loss} is 
\begin{equation}
\mathcal{L} = \E_{\boldsymbol{z}\sim p \left(\boldsymbol{z}\right)}\left[  \ln p(\boldsymbol{z}) - \ln \left|\det\left(\frac{\partial g(\boldsymbol{z}) }{\partial \boldsymbol{z}}\right)\right|  - \ln {{\pi}(g(\boldsymbol{z})) }\right], 
%\mathcal{L} = \E_{\boldsymbol{z}\sim p \left(\boldsymbol{z}\right)}\left[ E(g(\boldsymbol{z})) + \ln p(\boldsymbol{z}) - \ln \left|\det\left(\frac{\partial g(\boldsymbol{z}) }{\partial \boldsymbol{z}}\right)\right|  \right].
\label{eq:abatchofloss}
\end{equation}
where the log-Jacobian determinant can be efficiently computed by summing the contributions of each bijector. Notice that in \Eq{eq:abatchofloss} all the network parameters are inside the expectation,
%but not in the sampling process
which amounts to the \emph{reparametrization trick}~\cite{Goodfellow2016}. We perform stochastic optimization of \Eq{eq:abatchofloss}~\cite{Kingma2015}, in which the gradients with respect to the model parameters are computed efficiently using backpropagation. The gradient of \Eq{eq:abatchofloss} is the same as the one of the KL-divergence \Eq{eq:lowerbound} since the intractable partition function $Z$ is independent of the model parameter.  
%Note here the loss function is different from the typical ones used in the density estimation, where one assumes to already have access independent samples from the model. The learning only requires bare energy function the physical problem but not data which relies on efficient ways to generate sample according to the energy function.
%The loss function \Eq{eq:abatchofloss} is somewhat related to the supervised learning approach~\cite{Huang2017a, Liu2017a} except the training samples are now drew from the model probability density. 
%Using the loss function \Eq{eq:abatchofloss} the network is learning from samples generated by itself. 

Since the KL-divergence is asymmetric, the PDD is different from the Maximum Likelihood Estimation (MLE) which amounts to minimizing the empirical approximation of the KL-divergence in an \emph{opposite direction} $\KL*{\frac{\pi (\boldsymbol{x})}{Z}}{q(\boldsymbol{x})}$~\cite{Bishop2006, Goodfellow2016}. 
%Huang2017a, Liu2017a
The most significant difference is that in PDD one does not rely on an additional way (such as efficient MCMC) to collect independent and identically distributed configurations of the physical problem for training.  
%In principle, one can train it using the standard maximum likelihood estimation (MLE)~\cite{Goodfellow2016} on a dataset of physical configurations.
Moreover, optimizing the variational objectivity \Eq{eq:abatchofloss} can be more efficient than MLE because one directly makes use of the analytical functional form and gradient information of the target density $\pi(\boldsymbol{x})$. Finally, in the variational calculation, it is always better to achieve a lower value of the training loss \Eq{eq:abatchofloss} without the concern of overfitting~\cite{SM}.

The variational approach can also be integrated seamlessly with the MCMC sampling to produce unbiased physical results with enhanced efficiency. The partition function of the physical problem can be expressed in terms of the latent variables
\begin{equation}
Z =  \int \mathrm{d} \boldsymbol{z}\, \pi(g(\boldsymbol{z})) \left|\det\left( \frac{\partial g(\boldsymbol{z})}{\partial \boldsymbol{z}} \right) \right|=  \int \mathrm{d} \boldsymbol{z}\, p(\boldsymbol{z}) \left[\frac{\pi(g(\boldsymbol{z}))}{q(g (\boldsymbol{z}))} \right], 
\label{eq:change-of-variables}
\end{equation}
where the first equality simply invokes \emph{change-of-variables} from the physical space $\boldsymbol{x}$ to the latent space $\boldsymbol{z}$ using the learned normalizing flow, and the second equality rearranges terms using \Eq{eq:probflow}.

The integrand of \Eq{eq:change-of-variables} offers direct access to the renormalized energy function in the latent space induced by the flow $\boldsymbol{z}=g^{-1}(\boldsymbol{x})$. One sees that when the model density $q(\boldsymbol{x})$ perfectly matches the target density $\pi(\boldsymbol{x})/Z$, the energy function of the latent variables reduces to one associated with the prior $p(\boldsymbol{z})$. The variational calculation \Eq{eq:lowerbound} would then always push the latent distribution towards the independent Gaussian prior. Therefore, it would be advantageous to perform Metropolis~\cite{Metropolis1953} or hybrid Monte Carlo (HMC) sampling~\cite{Duane1987} in the latent space for better mixing. Given samples in the latent space, one can obtain the corresponding physical variable via $\boldsymbol{x} = g(\boldsymbol{z})$. This generalizes the Monte Carlo updates in the wavelet basis~\cite{ismail2003multiresolution1, ismail2003multiresolution2} to the case of  adaptively latent space for a given physical problem.

\begin{figure}[t]
\centering
\includegraphics[width=\columnwidth]{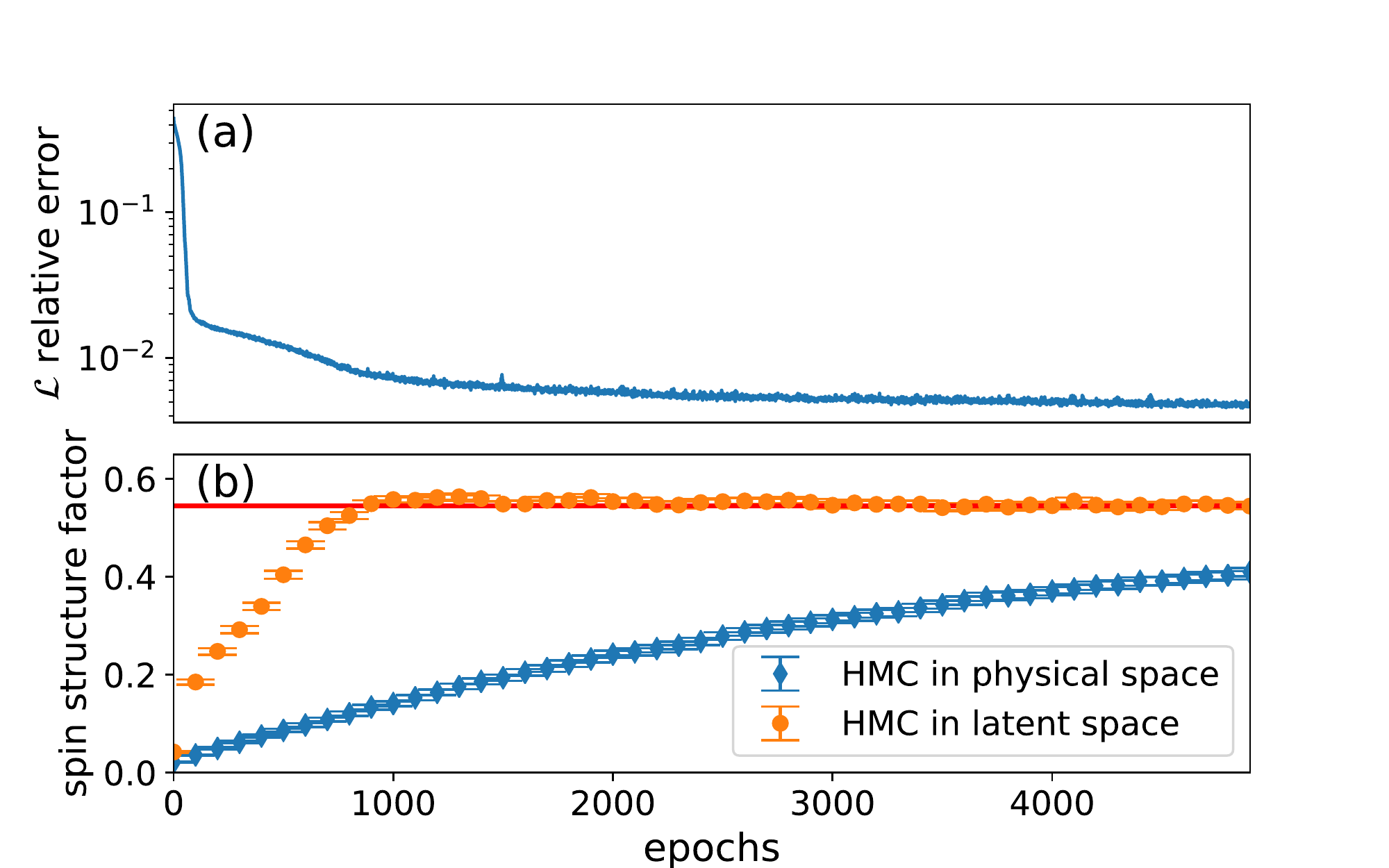}
\caption{Physical results obtained for the continuous field theory of ~\Eq{eq:contIsing} equivalent to the Ising model on a $N=16\times16$ lattice at critical coupling.   
(a) The relative error in the variational free energy \Eq{eq:loss} decreases with training epochs. The exact free energy is obtained from the analytical solution of the Ising model~\cite{Onsager1944, kaufman_crystal_1949}. (b) Uniform spin structure factor computed using hybrid Monte Carlo sampling in the latent and the physical spaces respectively. The errorbars are computed using independent batch of samples. The solid red line is the result of $\E_{\boldsymbol{s}\sim \pi_\mathrm{Ising} (\boldsymbol{s})} \left[\sum_{i,j} s_i s_j/N^2\right]$ computed directly for the Ising model.
%An epoch means one optimization step of the network parameters.
}
\label{fig:loss_acc}
\end{figure}

As a demonstration, we apply NeuralRG to the two-dimensional Ising model, a prototypical model in statistical physics. To conform with the continuous requirement of the physical variables, we employ the continuous relaxations trick of Refs.~\cite{fisher1983critical, Zhang2012}. We first decouple the Ising spins using a Gaussian integral, then sum over the Ising spins to obtain a target probability density
\begin{equation}
\pi(\boldsymbol{x}) = \exp\left(-\frac{1}{2} \boldsymbol{x}^{T} \left(K+\alpha I\right)^{-1}\boldsymbol{x}\right) \times\prod_{i=1}^{N} \cosh\left(x_{i}\right), 
\label{eq:contIsing}
\end{equation}
%\begin{equation}
%E(\boldsymbol{x}) = \frac{1}{2} \boldsymbol{x}^{T} \left(K+\alpha I\right)^{-1}\boldsymbol{x} - \sum_{i=1}^{N}  \ln\cosh\left(x_{i}\right), 
%\label{eq:contIsing}
%\end{equation}
where $K$ is an $N\times N$ symmetric matrix, $I$ is an identity matrix, and $\alpha$ is a constant offset such that  $K+\alpha I$ is positive definite~\footnote{We choose $\alpha$ such that the lowest eigenvalue of $K+\alpha I$ equals to $0.1$}. For each of the configurations, one can directly sample the discrete Ising variables $\boldsymbol{s}=\{\pm 1\}^{\otimes N}$ according to $\pi(\boldsymbol{s}|\boldsymbol{x})=\prod_{i} (1+e^{-2 s_{i}x_{i}})^{-1}$. It is straightforward to verify that the marginal probability distribution
$
\int \mathrm{d}{\boldsymbol{x}}\, \pi(\boldsymbol{s}|\boldsymbol{x})\pi(\boldsymbol{x})  \propto \exp \left( \frac{1}{2} \boldsymbol{s}^{T} K \boldsymbol{s}\right)
\equiv \pi_\mathrm{Ising}(\boldsymbol{s}) $ restores the Boltzmann weight of the Ising model with the  coupling matrix $K$. 
%~\footnote{Note that \Eq{eq:contIsing} is not a unique. By diagonalizing $(K+\alpha I)$ one obtains a model where each continuous variable couples to one Fourier component of the original Ising spins~\cite{fisher1983critical, Zhang2012}. Performing NeuralRG on such problem would correspond to the momentum space RG.}. 
Therefore, Equation (\ref{eq:contIsing}) can be viewed as a dual version of the Ising model, in which the continuous variables $\boldsymbol{x}$ represent the field couple to the Ising spins. We choose $K$ to describe the two-dimensional critical Ising model on a square lattice critical with periodic boundary condition.  %To further enhance the power of the network, we apply three layers of disentangler blocks between every two decimator layers.

We train the NeuralRG network of the structure shown schematically in Fig.~\ref{fig:nflow}(a) 
%with two modifications. First, the layout of the blocks is shown in Fig.~\ref{fig:tebd-tree-2d}(c), where we use two layers of disentanglers between every other decimator~\cite{Kim2017}. 
where the bijectors are of the size $2\times2$, as shown in Fig.~\ref{fig:tebd-tree-2d}(d). The results in Fig.~\ref{fig:loss_acc}(a) shows that the variational free-energy continuously decreases during the training. In this case, the exact lower bound reads $-\ln Z = -\ln Z_\mathrm{Ising}-\frac{1}{2} \ln \det(K+\alpha I)  +\frac{N}{2}[\ln (2/\pi) -\alpha]$, where $Z_\mathrm{Ising}=\sum_{\boldsymbol{s}}\pi_\mathrm{Ising}(\boldsymbol{s})$ is known from the exact solution of the Ising model~\cite{Onsager1944} on the finite periodic lattice~\cite{kaufman_crystal_1949}. % or numerics such as Wang-Landau sampling~\cite{Wang2001} and tensor network algorithms~\cite{Levin2007,Gu2008,Gu2009,Xie2009,Evenbly2015a}. 
We show results obtained in a wider temperature range and generated samples in the supplementary material~\cite{SM}. 

To make use of the learned normalizing flow, we perform the hybrid Monte Carlo (HMC)~\cite{SM} sampling in the latent space in parallel to the training using the effective energy function \Eq{eq:change-of-variables}. The physical results quickly converge to the correct value indicated by the solid red line. %is from an alternative Monte Carlo simulate of the Ising model, which servers as an confirmation.
In comparison, the HMC simulation in the original physical space using \Eq{eq:contIsing} as the energy function fails to thermalize during the same HMC steps. %Although the two sampling approaches give rise to the same physical results for the spin structure factor shown in Fig.~\ref{fig:loss_acc}(b), 
Even taking into account the overhead of training and evaluating the neural network, sampling in the latent space is still significantly more efficient
%since one deals with nearly Gaussian distributed independent latent variables in the HMC sampling. 

%Physical interpretation of the learned variables
To reveal the physical meaning of the learned latent variables, we recall the wavelets interpretation of the RG~\cite{guy1999wavelets, Qi_Exact_2013, PhysRevLett.116.140403}. 
%It is also interesting to examine the normalizing flow in this perspective. 
In our context, if each bijector performs the same linear transformation, the network precisely implements the discrete wavelet transformation~\cite{PhysRevA.97.052314}. Using the wavelets language, the bijectors at each layer extract "smooth" and "detail" components of the input signal separately. And the bijectors in the next layer perform transformations only to these "smooth" components. %For example, consider the simplest Haar wavelet, it corresponds to a linear bijector which computes the mean and difference of the two input variable. Physically, collective variables corresponds to the domain of various sizes at different locations. 

\begin{figure}[t]
\centering
\includegraphics[width=\columnwidth]{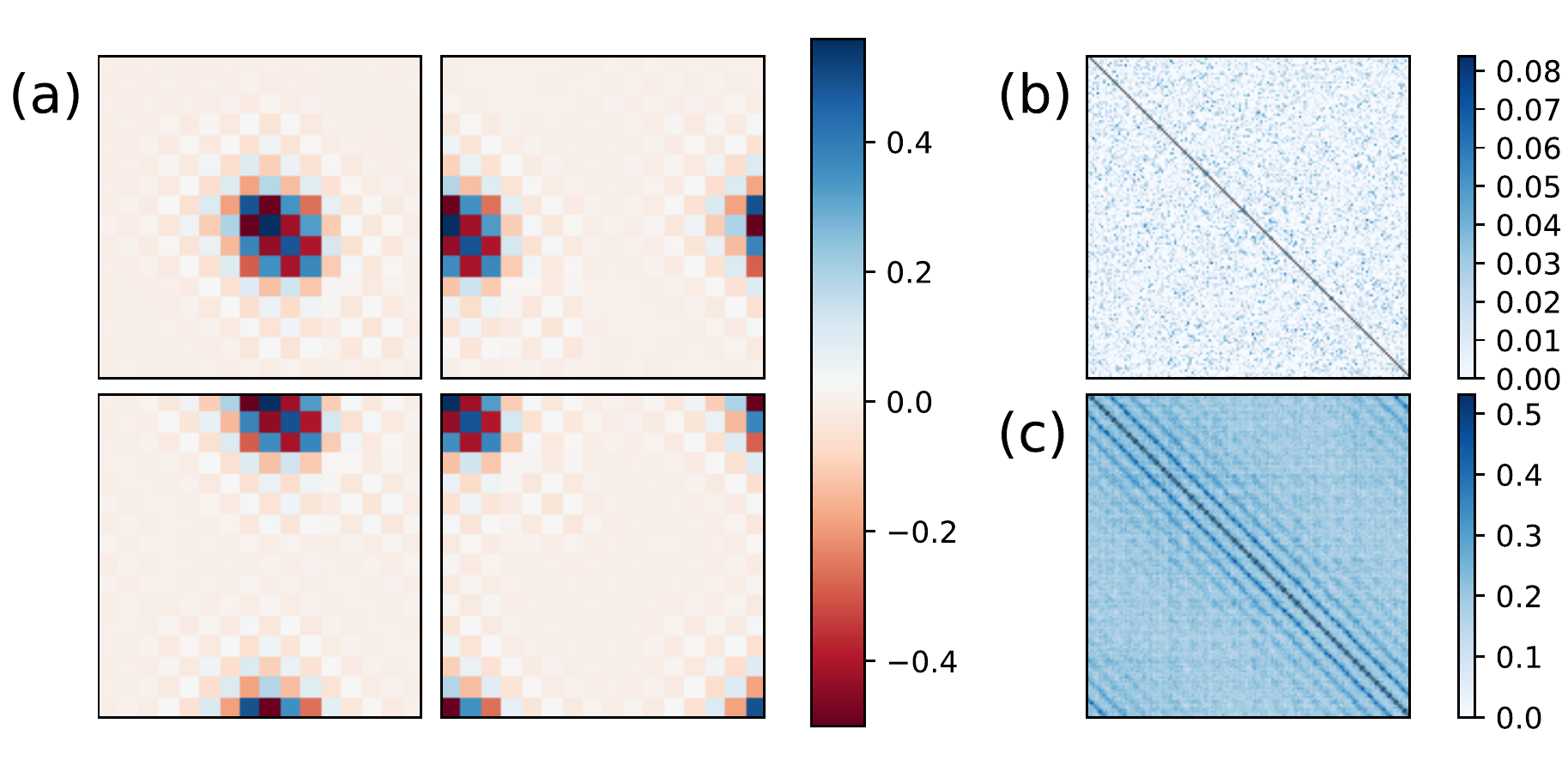}
\caption{(a) The responses of the latent space collective variables with respect to the physical variables $\E_{\boldsymbol{x}\sim \pi(\boldsymbol{x})} [\partial{z_i}/\partial{\boldsymbol{x}}]$. (b) Mutual information between the latent variables and (c) the physical variables. Note different scales in the colorbars of (b) and (c).}
\label{fig:visualize}
\end{figure}

We probe the response of the latent variables by computing the gradient of the transformation $\boldsymbol{z} = g^{-1}(\boldsymbol{x})$ using back-propagation through the network. Figure~\ref{fig:visualize}(a) visualizes the expected gradient $\E_{\boldsymbol{x}\sim \pi(\boldsymbol{x})} [\partial{z_i}/\partial{\boldsymbol{x}}]$ averaged over a batch of physical samples, where $z_{i}$ are the four top-level collective variables connecting to all of the physical variables. Each of them responds similarly to a nonoverlapping spatial region, which is indeed a reminiscence of the wavelets. On the other hand, the gradient $\partial{z_i}/\partial{\boldsymbol{x}}$ also exhibits variation for different physical variables. The variation is an indication of the \emph{nonlinearity} of the learned transformation since otherwise, the gradient is independent of data in the ordinary linear wavelets transformation. Thus, the latent variables can be regarded as a nonlinear and adaptive learned generalization of the wavelets representation. %A downside of the nonlinearities is that they hinder a clear interpretation of the physical meaning of the learned latent variables. 
Employing more advanced feature visualization and interpretability tools in deep learning~\cite{olah2017feature, olah2018the} may help distill more useful information from the trained neural network.   
%This is particularly problematic when the expressibility of the model is not sufficiently strong and target probability is multimodal. 

%MI
Finally, to characterize the effective interactions in the latent space, we plot estimated MI~\cite{PhysRevE.69.066138} between the latent variables in Fig.~\ref{fig:visualize}(b). 
The network does not map the physical distribution into ideally factorized Gaussian prior, in line with the gap to the exact free energy Fig.~\ref{fig:loss_acc}(a). However, the remaining MI between the latent variables is much smaller compared to the ones between the physical variables shown in Fig.~\ref{fig:visualize}(c). %Furthermore, we checked that the latent variables are less dependent even beyond the two-point correlations by estimating their mutual information. 
%As one can clearly see, the transformation removes the correlation between the original physical variables step by step. 
Obtaining a mutually independent representation of the original problem underlines the efficiency boost of the latent space HMC demonstrated in Fig.~\ref{fig:loss_acc}(b).
Adaptive learning of a \emph{nonlinear} transformation is a distinct feature of the present approach compared to \emph{linear} independent component analysis and wavelet transformations. These linear transformation approaches would not be able to remove dependence between the physical variables unless the physical problem is a free theory.  

The NeuralRG approach provides an automatic way to identify mutually independent collective variables~\cite{Barducci2011,Invernizzi2017}. Note that the identified collective variables do not need to be the same as the ones in the conventional RG. This significant difference is due to the conventional approach
focusing on identifying the fixed points under the iterative application of the same predetermined transformation to the
physical variables (e.g. block decimation or momentum shell integration). 
While the present approach aims at finding out a set of hierarchical transformations that map complex physical probability densities to the {predetermined prior distribution}. Thus, its application is particularly relevant to off-lattice  molecular simulations that involve a large number of continuous degrees of freedom which are often very difficult to simulate. %It is interesting to further study the renormalized collective variables from an information theory perspective.   
%Applying back to ML
%%%We focused on physical systems with translational invariance in this paper, where it makes sense to use a predetermined homogenous architecture. For physical systems with disorders~\cite{Ma1979,Fisher1992} or realistic dataset in machine learning, it would be interesting to learn the network structure based on the mutual information pattern of the problem~\cite{Chow1968,Hyatt2017}. 

%%%%Besides calling a revived attention to the probabilistic~\cite{Jona-Lasinio1975} and information theory~\cite{Apenko2012} perspectives on the RG flow, the NeuralRG framework also possesses mathematical properties underline modern understanding of RG in terms of diffeomorphism~\cite{lassig1990geometry, Caticha_Changes_2016}. 
Lastly, the conventional RG is a semigroup since the process is irreversible. However, the NeuralRG networks built on normalizing flows form a group, which can be useful for exploring the information preserving RG~\cite{PhysRevD.86.065007, Qi_Exact_2013, You2017} in conjunction with holographic mapping. %Taking the deep learning perspective~\cite{Goodfellow2016}, one tempts to view the RG as a continuous transformation of the data manifolds. However, since diffeomorphism function keeps the topology of the manifolds of physical and latent space, it is yet to be seen how does such assumption interplay with the topological features in statistical configurations. 

\begin{acknowledgments}
The authors thank Yang Qi, Yi-Zhuang You, Pan Zhang, Jin-Guo Liu, Lei-Han Tang, Chao Tang, Lu Yu, Long Zhang, Guang-Ming Zhang, and Ye-Hua Liu for discussions and encouragement. We thank Wei Tang for providing the exact free energy value of the 2D Ising model on finite lattices. The work is supported by the Ministry of Science and Technology of China under the Grant No. 2016YFA0300603 and the National Natural Science Foundation of China under Grant No. 11774398.
\end{acknowledgments}

\bibliography{RealNVP,refs}

\appendix

\clearpage
%%%%%%%%%% Merge with supplemental materials %%%%%%%%%%
\pagebreak
\widetext
\begin{center}
\textbf{\large Supplemental Materials: Neural Network Renormalization Group}
\end{center}
%%%%%%%%%% Merge with supplemental materials %%%%%%%%%%
%%%%%%%%%% Prefix a "S" to all equations, figures, tables and reset the counter %%%%%%%%%%
\setcounter{equation}{0}
\setcounter{figure}{0}
\setcounter{table}{0}
%\makeatletter
\renewcommand{\theequation}{S\arabic{equation}}
\renewcommand{\thefigure}{S\arabic{figure}}
%\renewcommand{\bibnumfmt}[1]{[S#1]}
%\renewcommand{\citenumfont}[1]{S#1}
%%%%%%%%%% Prefix a "S" to all equations, figures, tables and reset the counter %%%%%%%%%%
\numberwithin{equation}{section}
\section{Training Algorithm}

Algorithm \ref{alg:neuralrg} shows the variational training procedure of the NeuralRG network. %The experience buffer has a fixed maximum size. The distribution of the samples in the buffer approach to the target probability density during training. When building up the buffer, we perform data argumentation by appending inversion symmetry related configurations of the same weight. We prevent mode collapse using the NLL (\ref{eq:NLL}) evaluated on the samples drawn from the buffer. To initialize the Markov Chain \Eq{eq:acceptance} we also draw samples from the buffer. 

\begin{algorithm}[H]
	\begin{algorithmic}
		\caption{Variational training algorithm of the NeuralRG network}
	    \Require Normalized prior probability density \texttt{p(z)}, e.g. a Normal distribution
		\Require Unnormalized target probability density \texttt{pi(x)}
		\Ensure A normalizing flow neural network \texttt{x=g(z)} with normalized probability density \texttt{q(x)} 
	  \State Initialize a normalizing flow \texttt{g} 
	  \While{\texttt{Stop Criterion Not Met}} 
	     \State Sample a batch of latent variables \texttt{z} according to the prior \texttt{p(z)}
	     \State Obtain physical variables \texttt{x=g(z)} and compute their densities \texttt{q({x})} \Comment \Eq{eq:probflow}
	     \State \texttt{loss = mean\{ln[q(x)]-ln[pi(x)]\}} \Comment \Eq{eq:abatchofloss}
		\State Optimization step for the \texttt{loss}  	  
%		\State Physical space HMC with energy function \texttt{-ln[pi(x)]}
%	    \State Latent space HMC with energy function \texttt{-ln[p(z)]-ln[pi(g(z))]+ln[q(g(z))]}
	  \EndWhile		
	  \label{alg:neuralrg}
	\end{algorithmic}
\end{algorithm}

%\section{HMC sampling}
%Hybrid Monte Carlo approach is an efficient approach to sample from continuous distributions. It involves the gradient information of the probability density.

\section{Details about the Real NVP Bijector}

To implement the bijector we use the real-valued non-volume preserving (Real NVP) net~\cite{Dinh2016}, which belongs to a general class of bijective neural networks with tractable Jacobian determinant~\cite{Dinh2014, Germain2015, Rezende2015, Dinh2016, Kingma2016, Oord2016c, Oord2016a, Papamakarios2017,Diederik2018}. Real NVP is a generative model with explicit and tractable likelihood. One can efficiently evaluate the model probability density $q(\boldsymbol{x})$ for any sample, either given externally or generated by the network itself. %This feature is important for integrating with an unbiased Metropolis sampler. 

The Real NVP block divides the inputs into two groups $\boldsymbol{z} =\boldsymbol{z}_{<}\cup \boldsymbol{z}_{>}$, and updates only one of them with information from another group
%\begin{eqnarray}
%\boldsymbol{x}_{<} & = & \boldsymbol{z}_{<}, \nonumber \\ 
%\boldsymbol{x}_{>}  & =&  \boldsymbol{z}_{>} \odot e^{s(\boldsymbol{z}_{<})} + t(\boldsymbol{z}_{<}), 
%\label{eq:rnvp}
%\end{eqnarray}
\begin{equation}
\left\{
\begin{array}{ll}
\boldsymbol{x}_{<}  =  \boldsymbol{z}_{<},  \\ 
\boldsymbol{x}_{>}   =  \boldsymbol{z}_{>} \odot e^{s(\boldsymbol{z}_{<})} + t(\boldsymbol{z}_{<}), \label{eq:rnvp}
\end{array}
\right.
\end{equation}
where $\boldsymbol{x} =\boldsymbol{x}_{<}\cup \boldsymbol{x}_{>}$ is the output. $s(\cdot)$ and $t(\cdot)$ are two arbitrary functions parametrized by neural networks. In our implementation, we use multilayer perceptrons with $64$ hidden neurons of exponential linear activation~\cite{Clevert2015}. The output activation of the scaling $s$-function is a $\tanh$ function with learnable scale. While the output of the translation $t$-function is a linear function. The $\odot$ symbol denotes element-wise product. The transformation \Eq{eq:rnvp} is easy to invert by reversing the basic arithmetical operations. Moreover, the transformation has a triangular Jacobian matrix, whose determinant can be computed efficiently by summing over each component of the outputs of the scaling function $\ln \left|\det \left ( \frac{\partial \boldsymbol{x}}{\partial \boldsymbol{z}}\right )\right|=\sum_i [s(\boldsymbol{z}_<)]_i$. The transformation \Eq{eq:rnvp} can be composed by randomly sampling the bipartition so all variables are updated. In our implementation, we perform ten steps of the transformation \Eq{eq:rnvp} within each block. The log-Jacobian determinant of the bijector block is computed by summing up contributions of each layer. Within each layer on the same scale, we use the same block with shared parameters. The log-Jacobian determinant is computed by summing up contributions of each block.

%The RNVP block is not only invertible, and can also be done efficiently in the sense the two function evaluations have the same cost.
%One group of variables is used to update another group at each transformation step. 
%\begin{equation}
%y = b \odot x + (1-b) \odot \left(x \odot e^{s(b\odot x)} + t(b\odot x) \right), 
%\end{equation}

%\section{Toy Problems}
%We start by considering toy models in 2 dimension. The training samples were generated by via running the Metropolis independent sampler with $s\equiv0, t\equiv0$, which means that we generate the samples with a Gaussian prior. Next we train the real NVP network using both supervised and unsupervised approach and used the trained network to generate proposals. Figure~\ref{fig:correlation} shows the log-probability of the target data and the predictions of the model. One can clearly see that in both training approaches the network captures the important feature of the distribution. Next, we draw samples from the trained network. Figure~\ref{fig:proposal} shows that the proposed samples resembles the original distribution. Using such a proposal one can greatly enhances the acceptance ratio compared to using the Gaussian prior. 
%
%Figure .. shows that as the training goes the loss function decrease and acceptance rate continuously grows. As a consequence, the autocorrelations reduces. 

\section{Hybrid Monte Carlo in the latent space}
Hybrid Monte Carlo (HMC)~\cite{Duane1987} is a powerful sampling approach widely adopted in physics and machine learning~\cite{neal2011mcmc}. HMC reduces the diffusive behavior of the traditional Metropolis updates~\cite{Metropolis1953} via exploiting the Hamiltonian dynamics of continuous variables. Further acceleration of the HMC using neural networks is an active research direction in deep learning~\cite{Song2017, Levy2017}. 

For our application, we can either perform the HMC sampling of the physical variables given the normalized probability distribution or in the latent space. In the latter case, a key step of the HMC is the integration of the equation-of-motion according to effective energy function \Eq{eq:change-of-variables}. Note that the auto-differentiation tool in deep learning package conveniently provides tools to compute the force, i.e., the gradient of the energy with respect to the variables. Algorithm \ref{alg:hmc} outlines the key steps of the HMC simulation. 

\begin{algorithm}[H] 
	\begin{algorithmic}	  
      \caption{Hybrid Monte Carlo simulation in the latent space}
		\Require Energy function of the latent variables \texttt{U(z)=-ln[p(z)]-ln[pi(g(z))]+ln[q(g(z))]}  \Comment \Eq{eq:change-of-variables}
		\State Initial state of the latent variable \texttt{z}
	    \While{\texttt{Stop Criterion Not Met}} 
	     \State Sample velocity \texttt{v} from a Normal distribution
	     \State Leapfrog integration using the energy function \texttt{U(z)}
	     \State Metropolis acceptance according to the change of total energy \texttt{v$^T$v/2+U(z)} 
	    \EndWhile
	    \State  Obtain the physical variable \texttt{x=g(z)} and estimate physical observables
		\label{alg:hmc}
	\end{algorithmic}
\end{algorithm}

\section{Symmetrized variational calculation~\label{sec:symmetrize}}
One can further incorporate physical symmetries of the target problem into the variational scheme. As a concrete example, we discuss the implementation for the discrete inversion symmetry of the problem $\pi(\boldsymbol{x}) = \pi(-\boldsymbol{x})$. For general discussions, please refer to Ref.~\cite{moore2016symmetrized}. 

We introduce the symmetrized variational density 
\begin{equation}
  q_\mathrm{sym}(\boldsymbol{x}) = \frac{1}{2}\left[q(\boldsymbol{x}) + q(-\boldsymbol{x}) \right], 
  \label{eq:qsym}
\end{equation}
where $q(\boldsymbol{x})$ is given by the normalizing flow network. To evaluate $q(-\boldsymbol{x})$ we will first need to compute $\boldsymbol{z} = g^{-1}(-\boldsymbol{x})$, and then use \Eq{eq:probflow} of the main text. The density $q_\mathrm{sym}$ defined in this way manifestly respect the inversion symmetry. The training loss of symmetrized model reads 
\begin{align}
  \mathcal{L}_\mathrm{sym} & = \int{\mathrm{d}{\boldsymbol{x}}\, q_{\mathrm{sym}} (\boldsymbol{x}) \left[\ln{q_\mathrm{sym}(\boldsymbol{x})-\ln{\pi(\boldsymbol{x})}}\right]} \\  
  & = \int{\mathrm{d}{\boldsymbol{x}}\, q (\boldsymbol{x}) \left[\ln{q_\mathrm{sym}(\boldsymbol{x})}-\ln{{\pi(\boldsymbol{x})}}\right]}, 
\end{align}
where for the second equality we used the fact the expression in the square bracket is symmetric respect to inversion. Thus, the practical overhead of symmetrized calculation is merely evaluating $q_\mathrm{sym}(\boldsymbol{x})$ instead of $q(\boldsymbol{x})$. Compared to~\Eq{eq:loss} in the main text, the loss of the symmetrized model would always be lower since $ \mathcal{L}-  \mathcal{L}_\mathrm{sym}  = 
% = \int{\mathrm{d}{\boldsymbol{x}}q (\boldsymbol{x}) \left[\ln{q_\mathrm{sym}(\boldsymbol{x})-\ln{q(\boldsymbol{x})}}\right] } = 
\KL*{q(\boldsymbol{x})}{q_\mathrm{sym}(\boldsymbol{x})} \ge 0
$. Therefore, one can achieve better variational free energy by exploiting the  symmetry of the physical problem.

Writing the symmetrized density \Eq{eq:qsym} as a mixture model $q_\mathrm{sym}(\boldsymbol{x})= \frac{1}{2}\sum_{\eta = \pm }{q (\eta \boldsymbol{x})}$, we can treat  the sign variable $\eta$ on the equal footing with the latent variable $\boldsymbol{z}$.  
%The unsymmetrized probability $p(\boldsymbol{x}|\boldsymbol{z})$ is determinantal and can be wrote as
%\begin{equation}
%  p(\boldsymbol{x}|\boldsymbol{z}) = \delta(\boldsymbol{x}-g(\boldsymbol{z}))
%  \label{eq:unsymprob}  
%\end{equation}
%Then if we add symmetry transformation $\eta$ to it, this probability becomes
%
%where $g_{\eta}(\boldsymbol{x})$ is the same generate process as mentioned above but transfered according to symmetry $\eta$:
%\begin{equation}
%  g_{\eta}(\boldsymbol{z}) = T_{\eta}(g(\boldsymbol{z}))
%  \label{eq:gentran}  
%\end{equation}
In this regards, the generation process is deterministic given both $\boldsymbol{z}$ and $\eta$, i.e. $p(\boldsymbol{x}|\boldsymbol{z},\eta) = \delta(\boldsymbol{x}-\eta g( \boldsymbol{z})) 
$. Further marginalizing over the random sign $\eta$, one obtains the conditional probability $
  p(\boldsymbol{x}|\boldsymbol{z})=\frac{1}{2} \sum_{\eta}{\delta(\boldsymbol{x}-\eta g( \boldsymbol{z}))}$, which amounts to randomly flip the sign of the outcome of the normalizing flow network. Lastly, marginalizing over $\boldsymbol{z}$ in the joint probability $p(\boldsymbol{x}, \boldsymbol{z})=p(\boldsymbol{x}| \boldsymbol{z}) p(\boldsymbol{z}) $, one obtains the symmetric density \Eq{eq:qsym} as a consistency check.
%\begin{equation}
% q_\mathrm{sym}(\boldsymbol{x})= \frac{1}{2}\sum_{\eta = \pm }{q (\eta \boldsymbol{x})}
%\end{equation}
%where $q_{\eta}(\boldsymbol{x})$, as \Eq{eq:gentran}, can be written as
%\begin{equation}
%  q_{\eta}(\boldsymbol{x}) = q(T_{\eta} \boldsymbol{x})
%\end{equation}
 
For inferencing the latent variable given the physical variable, we compute the posterior using the Bayes' rule, 
\begin{equation}
p(\boldsymbol{z}|\boldsymbol{x}) 
= \frac{p(\boldsymbol{x}| \boldsymbol{z}) p(\boldsymbol{z})}{q_\mathrm{sym}(\boldsymbol{x})}
= \frac{1}{2 q_\mathrm{sym}(\boldsymbol{x})}\sum_{\eta=\pm}{ q(\eta\boldsymbol{x}) \delta(\boldsymbol{z}-g^{-1}(\eta\boldsymbol{x}))}. 
\end{equation}
Note that the two choices of the sign are weighted by $q(\eta\boldsymbol{x})$ in the posterior. In Fig.~\ref{fig:visualize}(a) the latent vector is inferred in this way, and the physical variables are flipped accordingly. Finally, the posterior also allows us to transform the physical probability density in the latent space 
\begin{align}
 \int{\mathrm{d} {\boldsymbol{x}}\, p(\boldsymbol{z}|\boldsymbol{x})  \pi(\boldsymbol{x}) } & = \frac{1}{2}\sum_{\eta=\pm} p(\boldsymbol{z}) \left [  \frac{ \pi({\eta g(\boldsymbol{z})}) }{q_\mathrm{sym}(\eta g(\boldsymbol{z}))}\right]
 \\ 
  & =    p(\boldsymbol{z}) \left [ \frac{ \pi(g(\boldsymbol{z}))}{q_\mathrm{sym}(g(\boldsymbol{z}))} \right ]. 
\end{align}
In the last equality, we used the symmetry condition of the target and the model densities. The resulting probability density is the same as \Eq{eq:change-of-variables} in the main text. 

\section{Additional results for a wider temperature range}

To gain a deeper understanding of the NeuralRG framework, we provide more data in a wider temperature range from the network of different depths. For better interpretability of the results, we use a single neural network which does not employ the $\mathbb{Z}_2$ symmetrized variational calculation~Appendix.~\ref{sec:symmetrize}. To suppress the model's tendency of collapsing into a single ferromagnetic solution, we add symmetry regularization term $\KL*{q(\boldsymbol{x})}{q(-\boldsymbol{x})}$ to the loss function \Eq{eq:loss}. 

\begin{figure}[h]
\centerline{\includegraphics[width=0.5\linewidth]{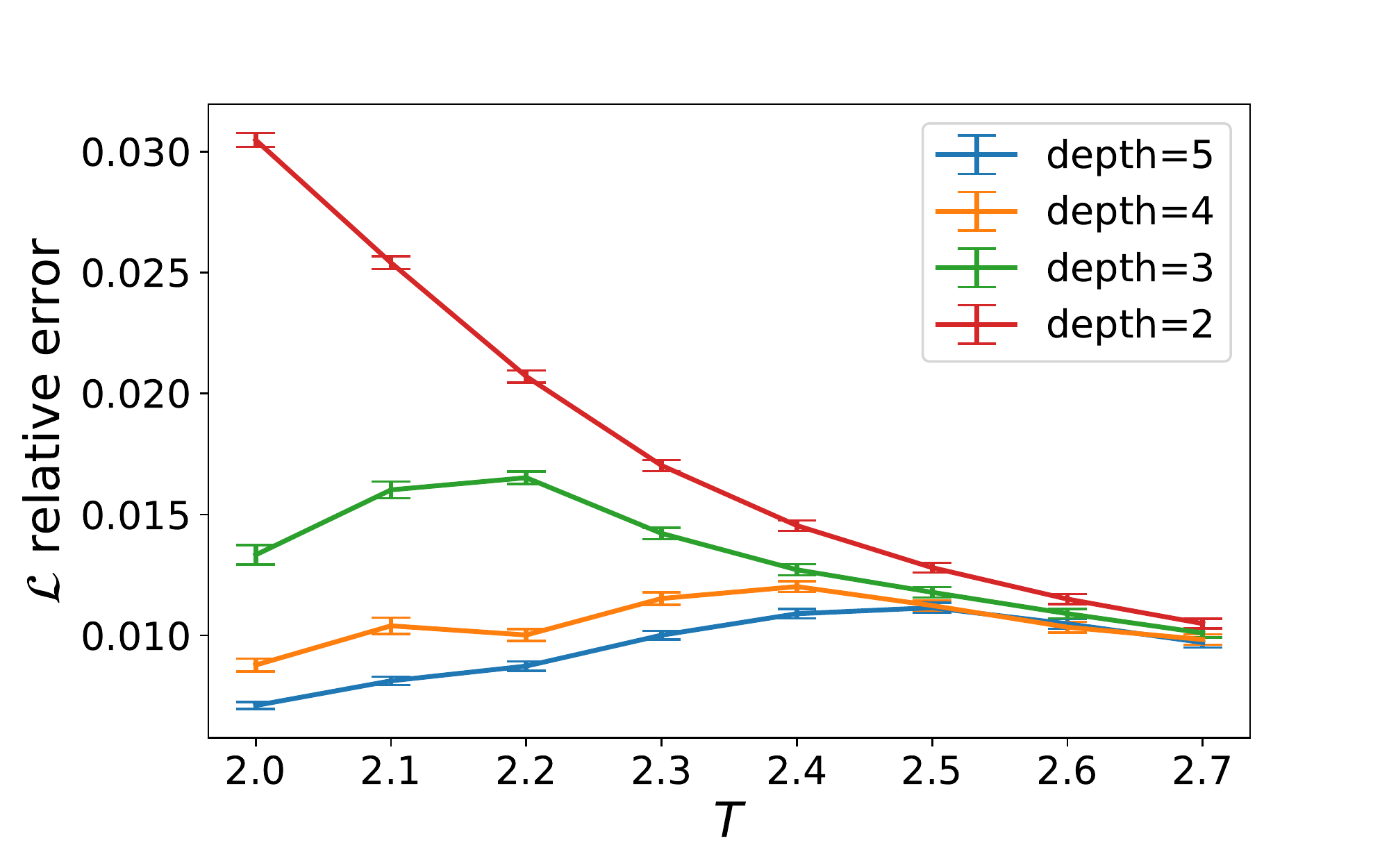}}
\caption{The relative error of the variational free energy versus temperature for networks of various depths. The deepest network has $5$ layers for the $N=32^{2}$ Ising model under consideration. And we construct shallower models by removing the top layers one by one. So the shallow network has a trapezoid shape. The errorbar is estimated from the variational loss computed on a batch of generated samples.
 }
\label{fig:scanT}
\end{figure}

Figure~\ref{fig:scanT} shows the relative error in the free energy across a wider temperature range. 
Deeper models consistently give lower variational free energy. This is expected since deep networks contain more variational parameters. 
The error of the variational free energy in general exhibit nonmonotonic behaviors as a function of temperature. This is because it is hard to capture critical fluctuations near the phase transition. While the network can simply produce nearly ferromagnetic configurations at low temperature, and short-range correlated configurations at high temperature. 
More importantly, shallow models perform almost as well as deep models at high temperature. This is due to that a few steps of transformations already suffice to remove short-range correlations from the physical distribution. While the advantage of deeper layers only show up at lower temperatures, and in particular, around the critical point, where the correlation length diverges. %Figure~\ref{fig:scanT}(b) shows the variance of the variational free energy, which shows similar behavior for temperature and depth of the network.  

\begin{figure}[tb]
\centerline{\includegraphics[width=0.5\linewidth]{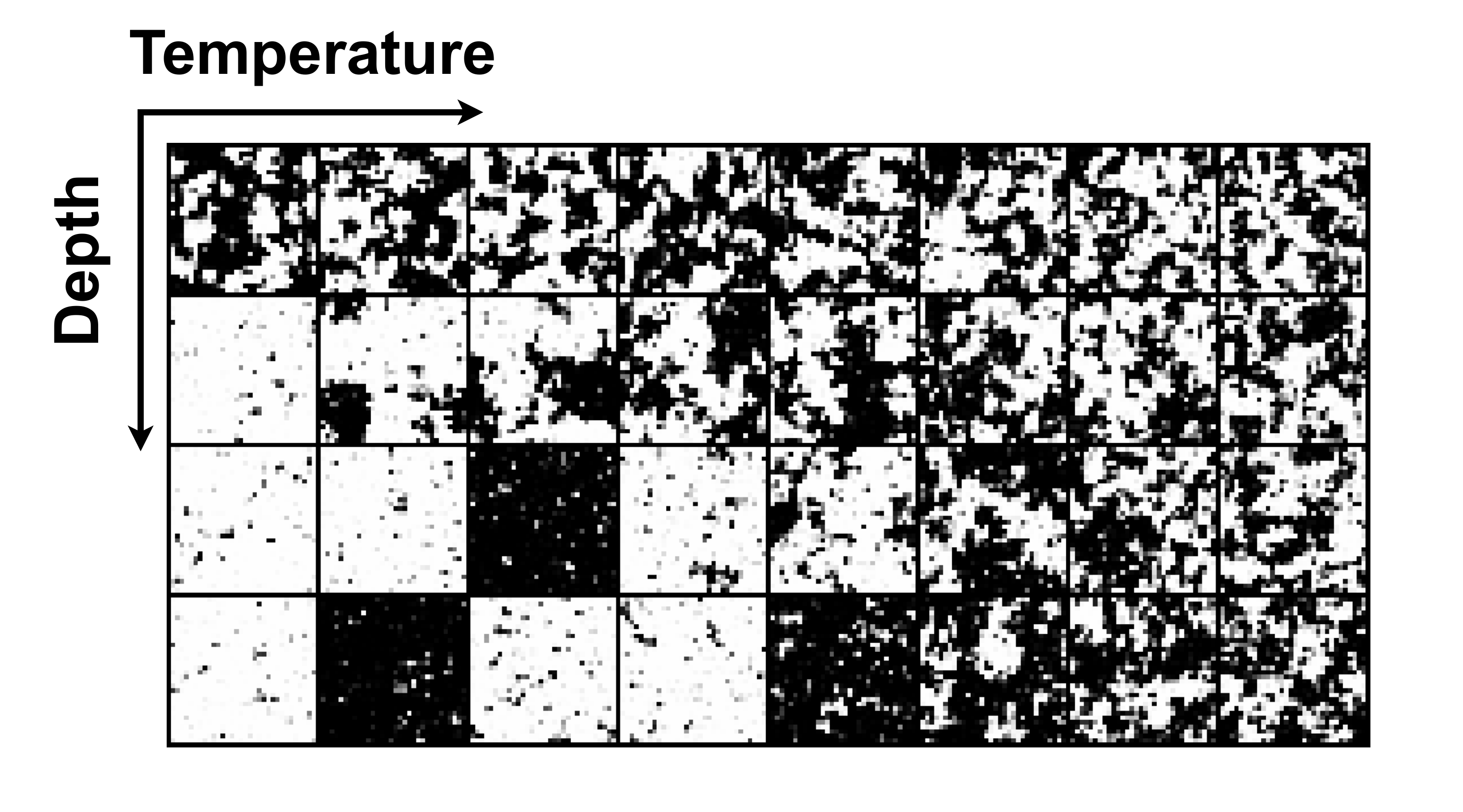}}
\caption{Directly generated samples from neural networks of different depth at various temperatures. 
From top to down, network depth $2,3,4,5$. From left to right, temperature $T=2.0,2.1,\ldots,2.7$.}
\label{fig:samples}
\end{figure}
To further gain intuitive understandings of these discussions, we plotted samples generated from the network in Fig.~\ref{fig:samples}. 
Shallow models cannot capture long-range correlation since during the generation process independent Gaussian variables do not pass through the same transformation block, and cannot be coupled together, c.f. Fig.~\ref{fig:tebd-tree-2d} of the main text.
For example, the upper left corner of Fig.~\ref{fig:samples} shows that the shallow network generates configuration with sharp domains at low temperature, which is not optimal in the free energy. In contrast, deeper models shown in the lower left corner better capture long-range coherence of physical variables. While moving to the right of Fig.~\ref{fig:samples}, at high temperatures all networks generate similar configurations since short-range correlations dominant at these temperatures.

\end{document}